\documentclass{IEEEtran}
\usepackage{pst-text}
\usepackage{epsfig}
\usepackage{url}
\usepackage[normalem]{ulem}
\usepackage{pstricks}
\usepackage{pst-node}
\usepackage{pst-rel-points}
\usepackage{cite} 
\usepackage{schemepgm}

\newcommand{\citeNP}{\cite}
\newcommand{\figtwo}{figure*}
\newcommand{\capsize}{}

\newcommand{\figrule}{\rule{\columnwidth}{.1mm}}
\newcommand{\figtworule}{\rule{\textwidth}{.1mm}}

\newenvironment{acks}{\section*{Acknowledgment}}

\newcommand{\deltarch}{\mbox{%
 \ensuremath{\delta_{\mbox{\footnotesize\sf rch}}}}}
\newcommand{\deltagc}{\mbox{%
 \ensuremath{\delta_{\mbox{\footnotesize\sf gc}}}}}
\newcommand{\GCstructure}{{\tt GC\_structure}}
\newcommand{\GCflag}{{\tt GC\_flag}}
\newcommand{\CreateTime}{{\tt Create\_time}}
\newcommand{\UseTime}{{\tt Use\_time}}

\newcommand{\cmt}[1]{}
\newcommand{\manual}[1]{{\protect#1}}

\newcommand{\silex}{\mbox{\tt silex}}
\newcommand{\lalr}{\mbox{\tt lalr}}
\newcommand{\eopl}{\mbox{\tt eopl}}
\newcommand{\prolog}{\mbox{\tt prolog}}
\newcommand{\sudoku}{\mbox{\tt sudoku}}
\newcommand{\cipher}{\mbox{\tt cipher}}

\newcommand{\TwoCells}[2]{%
\psset{unit=.25mm}
\begin{pspicture}(0,-2)(36,18)
\psframe(0,-5)(36,15)
\psline(18,-4)(18,15)
\putnode{z}{origin}{9}{5}{\rnode{#1}{}}
\putnode{z}{origin}{27}{5}{\rnode{#2}{}}
\end{pspicture}%
}

\newcommand{\xx}{{\tt x}}
\newcommand{\yy}{{\tt y}}

\begin{document}

\pagenumbering{arabic}

\title{Effectiveness of Garbage Collection in MIT/GNU Scheme}

\author{{Amey Karkare$^\ast$\thanks{$^\ast$Supported by Infosys Technologies Limited,
      Bangalore, under Infosys Fellowship Award.},
    Amitabha Sanyal and Uday Khedker}\\
  {Department of Computer Science \& Engg., \\
    IIT Bombay, Mumbai, India \\
    {\tt\{karkare,as,uday\}@cse.iitb.ac.in}}
}

\maketitle

\begin{abstract}
Scheme uses garbage collection for heap memory management.  Ideally,
garbage collectors should be able to reclaim all {\em dead} objects,
i.e. objects that will not be used in future. However, garbage
collectors collect only those dead objects that are not reachable
from any program variable.  Dead objects that are reachable from
program variables are not reclaimed.

In this paper we describe our experiments to measure the effectiveness
of garbage collection in MIT/GNU Scheme. We compute the drag time of
objects, i.e. the time for which an object remains in heap memory
after its last use.  The number of dead objects and the drag time
together indicate opportunities for improving garbage collection. Our
experiments reveal that up to \manual{26\%} of dead objects remain in
memory. The average drag time is up to \manual{37\%} of execution
time.  Overall, we observe memory saving potential ranging from
\manual{9\%} to \manual{65\%}.

\end{abstract}




\section{Introduction}
\label{sec:intro}
Garbage collection is an attractive alternative to manual memory
management because it frees the programmer from the responsibility of
keeping track of object lifetimes. This makes programs easier to
design, implement, understand and maintain. Ideally, a garbage
collector should be able to reclaim all {\em dead} objects,
i.e. objects that will not be used in future. However, this is not
possible because garbage collectors conservatively approximate the
{\em liveness} of an object by its reachability from a predefined set
of variables called {\em root} variables (typically the set of
variables on the program stack).  Garbage collectors cannot
distinguish between {live} reachable objects from {dead} reachable
objects. Hence they collect unreachable objects only as these objects
are guaranteed to be dead.  This means many dead objects are left
uncollected, a fact that has been confirmed by empirical studies for
various languages like Haskell~\cite{rojemo96lag} and
Java~\cite{shaham00gc,shaham01heap,shaham02estimating}.
Our experiments for MIT/GNU Scheme reveal that up to
\manual{26\%} of dead objects remain in memory, with dead objects
remaining in memory for up to \manual{37\%} of execution time.
Memory saving potential ranges from \manual{9\%} to \manual{65\%}.

\subsection{A Motivational Example}
\begin{figure}[ht]
\begin{center}
  \begin{tabular}{c}
    {\tt 
      {\begin{uprogram}
	\UFL\ (\LET\ ((x (\LIST\ \ldots)))
	\UNL{1} (\LET\ loop ((y x))
	\UNL{2}   (\IF\ (\NULLQ\ y)
	\UNL{3}         '()
	\UNL{3}         (\BEGIN
        \UNL{4}             (\ldots (\CAR\ y) \ldots)
	{{\rm\small;; process the head}}
	\UNL{4}             (loop (\CDR\ y))))))

      \end{uprogram}}
    } 
  \\ \\
  {\capsize (a) A program traversing a linked list.} \\ \\  \\
  \raisebox{-23mm}{\scalebox{.9}{
      \psset{unit=1mm}
      \psset{linewidth=.3mm}
      \begin{pspicture}(0,0)(48,45)
	\putnode{c}{origin}{10}{37}{{\white $O_1$} \TwoCells{c1}{c2} $O_1$}
	\putnode{d}{c}{10}{-10}{{\white $O_2$} \TwoCells{d1}{d2} $O_2$}
	\putnode{b}{c}{-10}{-10}{}
	\putnode{e}{d}{-13}{-12}{}
	\putnode{f}{d}{13}{-12}{{\white $O_3$} \TwoCells{f1}{f2} $O_3$}
	\ncline[nodesepB=-.5]{*->}{c2}{d}
	\ncline[nodesepB=-.5]{*->}{c1}{b}
	\ncline[nodesepB=2.5]{*->}{d1}{e}
	\ncline[nodesepB=-.5]{*->}{d2}{f}
	\putnode{w}{c}{-8}{8}{\psframebox[linestyle=none,framesep=.2]{{\xx}}}
	\ncline[nodesepB=-.2,angleA=330,angleB=120]{->}{w}{c}
	\putnode{i}{f}{-13}{-12}{}
	\putnode{j}{f}{13}{-12}{\psframebox[linestyle=none,framesep=.5]
	  {\rotatebox{-50}{\bf \ldots}}}
	\ncline[nodesepB=2.5]{*->}{f1}{i}
	\ncline[nodesepB=.1]{*->}{f2}{j}
	\putnode{y}{w}{28}{0}{\psframebox[linestyle=none,framesep=.2]{{\yy}}}
	\ncline{->}{y}{d}
      \end{pspicture}}} \\
  {\capsize (b) \renewcommand{\arraystretch}{.9}{
  \begin{tabular}[t]{@{}l}Memory graph at the beginning of\\
    the second iteration of the loop.\end{tabular}}}
  \end{tabular}
\end{center}
\caption{The Motivational Example.\label{fig:motiv}}
\figrule
\end{figure}
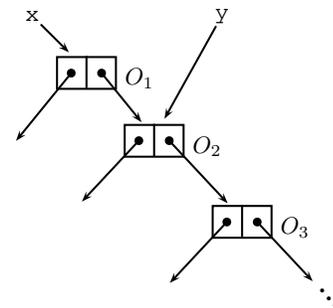

Figure~\ref{fig:motiv}(a)  shows a  program that  traverses  a singly
linked list.  Figure~\ref{fig:motiv}(b) shows  the memory graph at the
end of first iteration of the loop in the program. The object $O_1$ in
the memory is  unused after the first iteration  of the loop. However,
it cannot  be collected by garbage  collector as it  is reachable from
the  variable \xx.  Similarly,  object $O_2$  is  unused after  second
iteration of the loop, $O_3$ after third iteration, and so on.  All of
these objects, though  dead, will be garbage collected  only after the
variable \xx\ goes out of scope (i.e. at the end of the outer {\tt let}
loop.)
If \xx\  is nullified after its  last use (line 2,  first iteration of
loop)  the objects  may be  collected whenever  garbage  collection is
invoked after their last use, even though \xx\ remains in scope.

\subsection{Background}
\label{sec:background}
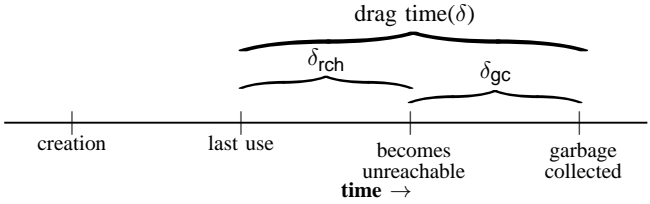
\begin{figure}[ht!]
  \begin{center}
\scalebox{.9}{
    \psset{unit=1mm}
  \begin{pspicture}(0,0)(110,31)
    \psrelpoint{origin}{st}{0}{13}
    \psrelpoint{st}{ct}{10}{0}
    \psrelpoint{ct}{fu}{10}{0}
    \psrelpoint{fu}{lu}{15}{0}
    \psrelpoint{lu}{ur}{25}{0}
    \psrelpoint{ur}{gc}{25}{0}
    \psrelpoint{gc}{end}{10}{0}
    \psline{-}(\x{st},\y{st})(\x{end},\y{end})
    \rput(\x{ct},\y{ct}){$|$}
    \rput(\x{lu},\y{lu}){$|$}
    \rput(\x{ur},\y{ur}){$|$}
    \rput(\x{gc},\y{gc}){$|$}
    \psrelpoint{ct}{lbct}{0}{-3}
    \rput(\x{lbct},\y{lbct}){\small creation}
    \psrelpoint{fu}{lbfu}{0}{-3}
    \psrelpoint{lu}{lblu}{0}{-3}
    \rput(\x{lblu},\y{lblu}){\small last use}
    \psrelpoint{ur}{lbur}{0}{-6}
    \rput(\x{lbur},\y{lbur}){\small
      \renewcommand{\arraystretch}{.9}{
	\begin{tabular}{c}becomes\\unreachable\end{tabular}}}
    \psrelpoint{gc}{lbgc}{0}{-6}
    \rput(\x{lbgc},\y{lbgc}){\small
      \renewcommand{\arraystretch}{.9}{
	\begin{tabular}{c}garbage\\collected\end{tabular}}}
    \psrelpoint{lu}{db0}{25}{14}
    \rput(\x{db0},\y{db0}){
      $\stackrel{
	\mbox{drag time($\delta$)}}{
	\mbox{\scalebox{8}[1.5]{$\overbrace{\mbox{}}$}}}$}

    \psrelpoint{lu}{db1}{12}{8}
    \rput(\x{db1},\y{db1}){
      $\stackrel{\deltarch}{
	\mbox{\scalebox{4}[1]{$\overbrace{\mbox{}}$}}}$}

    \psrelpoint{ur}{db0}{12}{6}
    \rput(\x{db0},\y{db0}){
      $\stackrel{\deltagc}{
	\mbox{\scalebox{4}[1]{$\overbrace{\mbox{}}$}}}$}
    \psrelpoint{st}{tm}{55}{-10}
    \rput(\x{tm},\y{tm}){\bf\small time $\rightarrow$}
  \end{pspicture}
  }
  \end{center}
  \caption{Events in the life of an object\label{fig:imp-eve}}
\figrule
\end{figure}
Figure~\ref{fig:imp-eve} shows  the important events in the  life of a
heap object: creation, use,  and garbage collection. The interval from
the time of  last use to the time of garbage  collection is called the
{\em  drag} time  ($\delta$) and  the object  is called  a  {\em dead}
object~\cite{rojemo96lag,shaham00gc,shaham01heap,shaham02estimating}.
If  an object  is never  used  after creation,  its drag  time is  the
interval between its  creation time and its collection  time.  A large
drag time indicates  that the object was reachable,  and hence ignored
by garbage collector, long after its last use.

The  number  of  dead objects  and  the  drag  time  is a  measure  of
improvement opportunities  in garbage collection and give  us an upper
bound on the  number of objects that could be  collected over the ones
collected by  garbage collector. The  upper bound is for  a particular
execution path of program. There may be no algorithm that can collect
all the dead objects.

The drag  time of an  object can be  divided into two  components: (1)
$\deltarch$, the interval  between the last use of  the object and the
time  when it becomes  unreachable, and  (2) $\deltagc$,  the interval
between the time when the object becomes unreachable and the time when
it is  collected by garbage  collector.  $\deltarch$ depends  upon the
program  but   is  independent  of  the   garbage  collector,  whereas
$\deltagc$  depends  heavily  upon the  garbage  collector---algorithm
used, frequency  of invocation, and time of  invocation.  For example,
for a  reference count based  garbage collector, $\deltagc$  is always
$0$  for  all  objects\footnote{Ignoring  objects  that  are  part  of
cycle.}. For  mark and sweep  or copying collector, $\deltagc$  for an
object depends upon  the time when the garbage  collector gets invoked
after  the object  becomes  unreachable. $\deltagc$  can typically  be
reduced  by increasing  the frequency  of garbage  collection  (at the
expense of slowing down the real computation).

\subsection{Organization}
\label{sec:organization}
The rest of this paper is organized as follows:
In Section~\ref{sec:experiments}, we describe our setup to carry out
the experiments and the benchmark programs used for measurements.
Section~\ref{sec:results} discusses the results of the experiments.
Section~\ref{sec:related}  describes the  research done  by  others in
related areas.
Section~\ref{sec:concl-and-future}  concludes the  paper  and provides
directions for future work.

\section{Experimental Setup}
\label{sec:experiments}
In our experiments  we measure the value of  $\deltarch$, which is the
characteristic  of the  program only  and independent  of  the garbage
collector,  and hence  independent of  the Scheme  implementation.  We
approximate $\deltarch$ by $\delta$ by forcing garbage collector to be
invoked  at a  very high  frequency, thereby  ensuring  that $\deltagc
\equiv 0$.  However, this technique  does not work for  incremental or
generational garbage collectors, because these do not scan all objects
in memory for every cycle.  Therefore, even after an object has become
unreachable, it may or may not be collected by next garbage collection
cycle, resulting in a non-zero $\deltagc$.

We   have  used  MIT/GNU   Scheme\footnote{Release   7.7.90.+,  from
\url{http://ftp.gnu.org/gnu/mit-scheme/snapshot.pkg/20050724/}}, as it
uses  a  simple copying  based  garbage  collector,  which is  neither
incremental  nor  generational.   It   is  also  easy  to  modify  the
implementation to invoke garbage collector at a very high frequency.

We record the statistics associated with {\em pairs} and {\em vectors}
only, ignoring all other  constructs (e.g.  {\em strings}) that create
objects  in  heap.   Collecting   statistics  for  all  constructs  is
difficult as (a)  it slows down the experiments  considerably, and (b)
the amount of statistics generated is overwhelming---even for moderate
size benchmarks, the execution goes out of memory. This restriction is
not that bad  because previous studies have shown  that cons cells and
vectors  account for  most  of the  space  as well  number of  objects
allocated                in                typical                LISP
programs~[\citeNP{zorn89phd},~Section~3.7.1].


We  associate a  structure with  every object  under  consideration to
record  the creation  time  and  the most  recent  use time.  Whenever
garbage collector  collects an  object, its data  is written to  a log
file  along  with the  garbage  collection  time.  The log  file  thus
generated         is         post-processed        to         generate
statistics.  Section~\ref{sec:gen-data} and Section~\ref{sec:rep-data}
describe the process in detail.

\subsection{Generating Data}
\label{sec:gen-data}
We associate  a structure (\GCstructure)  to record the  creation time
(\CreateTime)  and the  most  recent use  time  (\UseTime) with  every
object  under consideration.  The object  address is  used as  key for
\GCstructure.  \GCstructure\ also  contains a  flag (\GCflag)  to tell
whether the corresponding object  was collected by the current garbage
collection or  not. Scheme primitives  and procedures are  modified to
update the  fields of \GCstructure. We  describe how this  is done for
primitives that  operate on {\em  pairs} (or lists).   Similar changes
are applied to primitives for {\em vectors} too.
\begin{itemize}
\item{\em Creation:}  In Scheme,  pairs are created  using primitives,
  e.g.   {\tt   cons},    {\tt   list},   {\tt   vector->list},   {\tt
  string->list}.   These  primitives  are   modified  to   create  the
  {\GCstructure}(s)  corresponding  to the  new  pair(s) created,  and
  populate the \CreateTime, while \UseTime\ is set to an invalid value
  (-1).
  
\item{\em Use:} Primitives like  {\tt car}, {\tt cdr}, {\tt set-car!},
  {\tt set-cdr!}  including predicates like {\tt  null?}, {\tt pair?},
  {\tt  number?}  are  considered as  use  of their  argument and  are
  modified to  update \UseTime\  of corresponding \GCstructure.  If an
  object is never used, its \UseTime\ remains -1.

\item{\em Garbage collection:} Garbage  collector in MIT/GNU Scheme is
  a copying collector.  Before actual garbage collection, we reset the
  GC flag  in all {\GCstructure}s.  Whenever an object is  copied from
  working memory  to free memory, corresponding {\GCflag}  is set, and
  its  new address  is copied  into  the key.  At the  end of  garbage
  collection, all {\GCstructure}s are  scanned. If {\GCflag} is false,
  meaning the object was not copied to free memory, than the object is
  assumed to be collected by  garbage collector. For all such objects,
  we write  the data in {\GCstructure}s  to a log file  along with the
  garbage  collection  time. Also,  at  program  termination, data  in
  {\GCstructure}s of all  the objects remaining in heap  is written to
  the log file.
\end{itemize}
Scheme   runtime  libraries   are   forced  to   use  these   modified
primitives.  We   trigger  garbage  collection   at  every  \manual{10
milliseconds}. The benchmarks are  run in this modified environment to
generate data in the log file.

\subsection{Reporting Data}
\label{sec:rep-data}
The  log  files  generated  by  running  the  benchmark  programs  are
processed to generate statistics.  In our experiments
\begin{itemize}
\item  We  compare the  number  of dead  objects  with  the number  of
  allocated objects.
\item We  compare the  average drag time  of objects and  maximum drag
  time over all the objects with the total runtime of the program.
\item  We  record the  distribution  of drag  times  of  objects as  a
  percentage of total runtime.
\end{itemize}
Potential savings in  memory is estimated by measuring  the space time
product for dead objects as a percentage of space time product for all
allocated objects.

\subsection{Benchmarks}
\label{sec:benchmarks}
\begin{table}[t!]
\begin{center}
\renewcommand{\arraystretch}{1}{
\begin{tabular}{|l|l|} \hline
  \multicolumn{1}{|c}{\bf Benchmark} & \multicolumn{1}{|c|}{\bf
    Description} \\ \hline\hline
  \silex  & Lexical analyzer generator~\cite{silex.web} \\
  \lalr   & An LALR(1) parser generator~\cite{lalr.web}\\
  \eopl   & Code in Chapter 7 of Essentials of Programming \\
          & Languages~\cite{eopl.book} \\
  \prolog & Interpreter for pure Prolog~\cite{prolog.web}\\
  \sudoku & Sudoku~\cite{sudoku.www} puzzle solver~\cite{sudoku.web}\\
  \cipher & Program~\cite{cipher.web} to decode substitution cipher~\cite{cipher.www} \\ \hline
\end{tabular}}
\end{center}
\caption{The benchmarks}\label{tab:benchmarks}
\figrule
\end{table}

Our benchmark programs are described in Table~\ref{tab:benchmarks}.
The programs range from code (\eopl) from standard text-book {\em
Essentials of Programming Languages}~\cite{eopl.book} to the programs
(\cipher\ and \sudoku) by first year undergraduates.
\silex\ and \lalr\ are run with one test case each, while \eopl,
\prolog, \sudoku\ and \cipher\ are run with multiple test cases
each. The benchmarks and test cases can be obtained
from~\cite{gc-scheme.web}.

\section{Results}
\label{sec:results}

In this section we describe the results of our experiments.

\subsection{Reachable vs. Live Objects}
\label{sec:rch-use}
\begin{\figtwo}[t]
\begin{center}
  \begin{tabular}{|@{}c@{}|@{}c@{}|} \hline
    \scalebox{.99}{\includegraphics{BM_data/silex.epsi}}  &
    \scalebox{.99}{\includegraphics{BM_data/lalr.epsi}}  \\ \hline
    &\\
    \scalebox{.99}{\includegraphics{BM_data/eopl.epsi}}   &
    \scalebox{.99}{\includegraphics{BM_data/prolog.epsi}} \\ \hline
    &\\
    \scalebox{.99}{\includegraphics{BM_data/sudoku.epsi}} &
    \scalebox{.99}{\includegraphics{BM_data/cipher.epsi}} \\ \hline
  \end{tabular}
\end{center}

\hspace*{17mm}{\capsize  X axis  indicates  measurement instants  in
  milliseconds.  Y axis indicates  number of heap objects.} \\[-1mm]
\hspace*{17mm}{\capsize Solid line represents reachable objects.  Dashed
  line represents live objects.}

\caption{Reachable objects vs. live objects}
\label{fig:graph-rch-use}
\figtworule
\end{\figtwo}

Figure~\ref{fig:graph-rch-use} plots reachable objects and live
objects against time. The difference between the two lines gives the
number of reachable but dead objects. All the graphs show a
significant number of dead objects.

The graphs of \prolog, \sudoku\ and \cipher\ contain many {crests} and
{troughs},  while  the  graphs  for  \silex,  \lalr\  and  \eopl\  are
relatively  smooth. Our conjecture  is that  this is  because \prolog,
\sudoku\  and \cipher\  use backtracking  algorithms, and  the troughs
correspond   to  the   transitions  between   successive  backtracking
phases. To  validate our conjecture, we experimented  with \sudoku. We
used 3  different test cases---the first  test case had  only one cell
unfilled so that no backtracking  was required by \sudoku\ solver. The
second test case  had very few unfilled cells so  that a little amount
of backtracking  was involved,  while the third  test case was  a very
hard puzzle that involved high amount of backtracking. The results are
shown in  Figure~\ref{fig:sudoku-bts}: We can see that,  for the third
test case, the number of crest-trough pairs is too high as compared to
other  cases. Also,  since  \sudoku\ solver's  algorithm  is mainly  a
backtracking  algorithm with  a few  heuristics, runtime  of  the test
cases increases with the level of difficulty (backtracking).

\begin{\figtwo}[t]
\begin{center}
  \begin{tabular}{|@{}c@{}|@{}c@{}|@{}c@{}|} \hline
    & & \\
    \scalebox{.8}{\includegraphics{sudoku_nobt.epsi}}  &
    \scalebox{.8}{\includegraphics{sudoku_bt.epsi}}  &
    \scalebox{.8}{\includegraphics{sudoku_highbt.epsi}}   \\
    {\capsize No backtracking} & {\capsize Little backtracking} &
    {\capsize High backtracking} \\ \hline
  \end{tabular}
\end{center}

\hspace*{3mm}{\capsize  X axis  indicates  measurement instants  in
  milliseconds.  Y axis indicates  number of heap objects.} \\[-1mm]
\hspace*{3mm}{\capsize Solid line represents reachable objects.  Dashed
  line represents live objects.}

\caption{Effect of backtracking on \sudoku\ solver}
\label{fig:sudoku-bts}
\figtworule
\end{\figtwo}

In \eopl\ there is an initial burst where the reachable memory is very
high. This corresponds to the phase where all the test cases are
loaded into Scheme. For our experiments, we ran three interpreters
corresponding to the code given in Chapter~7
of~\cite{eopl.book}. Small crests in plot of reachable objects (and
troughs in plot of live objects) in \eopl\ correspond to transition
from one interpreter to other.

\begin{table}[t]
  \begin{center}
    \renewcommand{\arraystretch}{1}{
      \begin{tabular}{|l|r|r|r|} \hline
\multicolumn{1}{|@{\ }c@{\ }|}{\bf Benchmark} &
\multicolumn{1}{|@{\ }c@{\ }|}{\bf Reachable}  &
\multicolumn{1}{|@{\ }c@{\ }|}{\bf Live} &
\multicolumn{1}{|@{\ }c@{\ }|}{\bf Potential}\\
\multicolumn{1}{|@{\ }c@{\ }|}{} &
\multicolumn{1}{|@{\ }c@{\ }|}{\bf Object}  &
\multicolumn{1}{|@{\ }c@{\ }|}{\bf Object} &
\multicolumn{1}{|@{\ }c@{\ }|}{\bf Savings}\\
\multicolumn{1}{|@{\ }c@{\ }|}{} &
\multicolumn{1}{|@{\ }c@{\ }|}{\bf Integral}  &
\multicolumn{1}{|@{\ }c@{\ }|}{\bf Integral} &
\multicolumn{1}{|@{\ }c@{\ }|}{\bf \%}\\ \hline\hline
\silex & 409442730 & 141309450 & 65.48 \\
\lalr & 109380 & 58450 & 46.56 \\
\eopl & 373865300 & 217799490 & 41.74 \\
\prolog & 175096720 & 72172390 & 58.78 \\
\sudoku & 496456510 & 450879850 & 9.18 \\
\cipher & 208383570 & 184187520 & 11.61 \\
\hline
\end{tabular}
}
  \end{center}

  \caption{Space time product for reachable objects and
    live objects}
  \label{tab:space-time-prod}
\figrule
\end{table}

To estimate the memory savings, we compute the space-time product for
reachable objects and that for live objects by computing the area
under respective plots (see Table~\ref{tab:space-time-prod}).  The
potential of saving ranges from \manual{9\%} to \manual{65\%}
for our benchmarks.

\subsection{Number of  Allocated Objects vs. Dead Objects}
\label{sec:num-tot-drag}

\begin{\figtwo}[t]
\begin{center}
  \scalebox{.99}{\includegraphics{BM_data/TotVsDrag.epsi}}

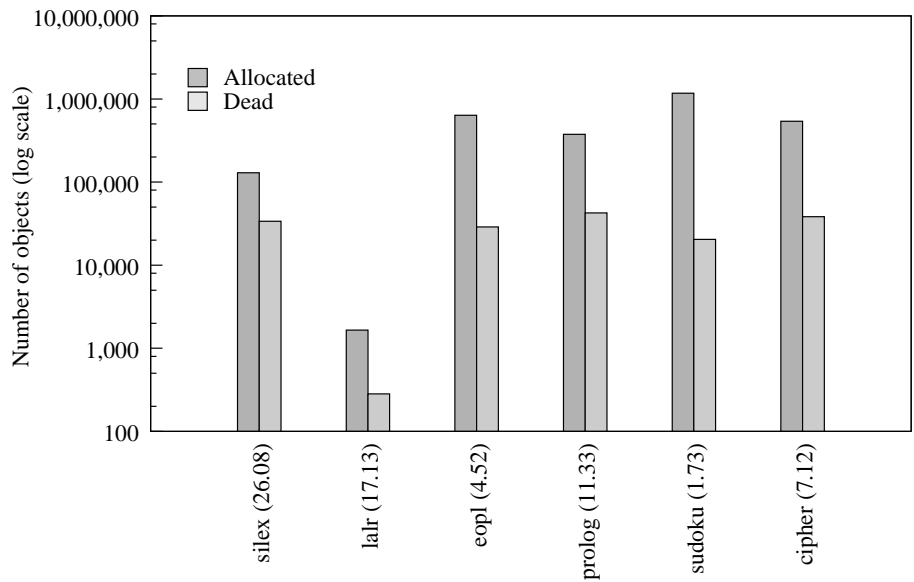
{\capsize Figures in parenthesis denote the percentage of dead objects
  with respect to allocated objects.}
\end{center}
\caption{Allocated objects vs. dead objects}
\label{fig:num-tot-drag}
\figtworule
\end{\figtwo}

Figure~\ref{fig:num-tot-drag} shows total number of objects allocated
vs. number of dead objects.  Even though the percentage of dead
objects is very small (less than \manual{5\%}) for \sudoku\ and \eopl,
there is still significant potential for memory savings because the
drag time of these objects is large. This is described in details in
next section.

\subsection{Drag vs. Runtime}
\label{sec:drag-run}

\begin{table}[t]
\begin{center}
    \begin{tabular}{|l|r|r|r|} \hline
\multicolumn{1}{|@{\ }c@{\ }|}{\bf Benchmark} &
\multicolumn{1}{|@{\ }c@{\ }|}{\bf Runtime}  &
\multicolumn{1}{|c|}{\bf Maximum Drag}  &
\multicolumn{1}{|c|}{\bf Average Drag}  \\ \hline\hline
\silex & 27950 & 27110 (96.99) & 7928.94 (28.36) \\
\lalr & 480 & 250 (52.08) & 179.96 (37.49) \\
\eopl & 109060 & 108620 (99.59) & 5403.56 (4.95) \\
\prolog & 39970 & 39700 (99.32) & 2419.81 (6.05) \\
\sudoku & 82730 & 82610 (99.85) & 2229.23 (2.69) \\
\cipher & 27250 & 13440 (49.32) & 630.25 (2.31) \\
\hline
\end{tabular}

\vskip 5pt
\end{center}
\hspace*{5mm}{ All times are in milliseconds.} \\
\hspace*{5mm}{ Figures in  parenthesis denote percentage value
  with respect to runtime.}

\caption{Statistics of dead objects}
\label{tab:drag-stats}
\figrule
\end{table}

In  Table~\ref{tab:drag-stats},  we  show  how drag  time  of  objects
compare with  the total  runtime for a  given program.  Note  that for
\silex, \eopl,  \prolog, and  \sudoku, the maximum  drag time  is very
close to  the total runtime.  This indicates presence of  objects that
are  created near  the  beginning  of the  program  and remain  unused
throughout the execution.

\begin{\figtwo}[t]
\begin{center}
  \begin{tabular}{|c|c|} \hline
    &\\
    \scalebox{.69}{\includegraphics{BM_data/silex_drag.epsi}}  &
    \scalebox{.69}{\includegraphics{BM_data/lalr_drag.epsi}}   \\ \hline
    &\\
    \scalebox{.69}{\includegraphics{BM_data/eopl_drag.epsi}}   &
    \scalebox{.69}{\includegraphics{BM_data/prolog_drag.epsi}} \\ \hline
    &\\
    \scalebox{.69}{\includegraphics{BM_data/sudoku_drag.epsi}} &
    \scalebox{.69}{\includegraphics{BM_data/cipher_drag.epsi}} \\ \hline
  \end{tabular}
\end{center}
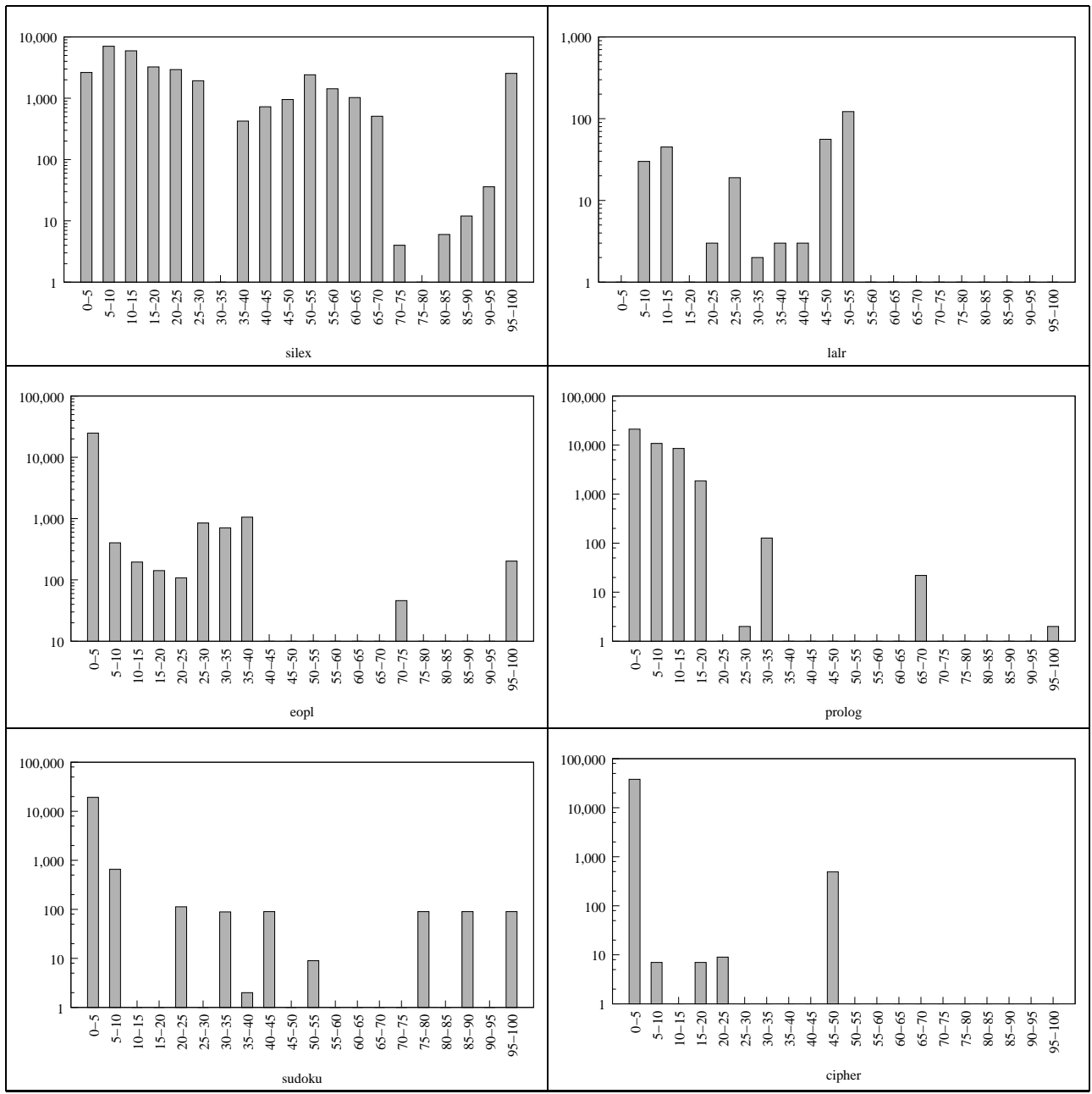
\hspace*{7mm}{\capsize X axis indicates percentage of runtime.  Y axis
  indicates number of dead heap objects (log scale).}

\caption{Distribution of dead objects as percentage of runtime}
\label{fig:graph-drag}
\figtworule
\end{\figtwo}

Figure~\ref{fig:graph-drag} shows the distribution of dead times of
objects as percentage of runtime of the program. In all the
benchmarks, most of the dead objects are in the range
\mbox{\manual{0--50\%}} of total runtime. \lalr\ and \cipher\ do not
have any objects towards higher percentages. On the other hand,
\silex, \eopl\ and \sudoku\ have a large number of objects that have a
significant drag time of \mbox{\manual{95--100\%}}. These objects
contribute significantly to the space time product
(Table~\ref{tab:space-time-prod}). Collecting such objects will yield
high memory savings.

\section{Related Work}
\label{sec:related}
Similar experiments have been done to measure the effectiveness of
garbage collection in different language implementations,
e.g. Haskell~\cite{rojemo96lag}, Java~\cite{shaham00gc,%
shaham01heap,shaham02estimating}.  Our definitions and
measurement methodologies are based on the standards from previous
work.

KBDB\cite{serrano00understanding}, a heap inspector for Scheme
programs, relies on user interaction to inspect heap usage at
different points during the execution of program. Typically, the heap
is inspected before and after evaluation of some expression to
estimate the memory leaked by that expression. This approach is
orthogonal to our approach of using dead object information to detect
memory leak.

\section{Conclusions and Future Work}
\label{sec:concl-and-future}
Our experiments  show that for Scheme,  at any given time,  there is a
significant number  of reachable objects  that are not live.   Also, a
large  number   of  such   objects  remain  in   memory  for   a  long
duration. Garbage  collection for Scheme can  improve significantly if
such objects  can be identified  and made unreachable at  the earliest
using automatic techniques.

In, our earlier work~\cite{khedker06heap}, we have shown that for
imperative languages like Java, the number of reachable dead objects
can be reduced by automatically identifying and nullifying dead memory
links. We are extending that work to be applicable to functional
languages. The work reported in this paper is the first step towards
that direction.
\begin{acks}
 We are thankful to Chris Hanson, Matt Birkholz and Taylor Campbell
for answering our queries related to memory management primitives in
MIT/GNU Scheme.
\end{acks}

\bibliography{scheme_gc}

\end{document}